\documentclass{article}
\usepackage{blindtext}
\usepackage{multicol}
\usepackage{geometry}
\usepackage{epsfig}
\usepackage{amssymb}
\usepackage{caption}
\usepackage{graphicx}
\usepackage{subfig}
\usepackage{siunitx}
 
\title{The DESY axion search program}
\author{D. Heuchel, A. Lindner, I. Oceano \\
Deutsches Elektronen-Synchrotron DESY, Notkestr. 85, 22607 Hamburg, Germany}
\date{February 14, 2023}
\begin{document}
\maketitle


\section{Introduction}
Feebly Interacting Particles (FIPs) might offer the solution to (some of) the open questions beyond the Standard Models of particle physics and cosmology.   
At DESY in Hamburg, three non-accelerator-based experiments will search for FIPs as dark matter candidates (ALPS~II, BabyIAXO) or constituting the dark matter in our home galaxy (MADMAX). Such experiments have to strive for sensitivities many orders beyond the reach of collider or beam-dump experiments. 
Among FIPs, the axion as motivated by the lack of any observed CP violation in Quantum Chromodynamics (QCD) \cite{Abel:2020pzs, Peccei:1977hh,Weinberg:1977ma,Wilczek:1977pj}, is frequently being used as a benchmark to compare the sensitivities of  experimental efforts.
Axions result from a new global Peccei-Quinn symmetry $U(1)$ that spontaneously breaks at the scale $f_a$.
\begin{figure}[!ht]
\begin{center}
\includegraphics[scale=0.5]{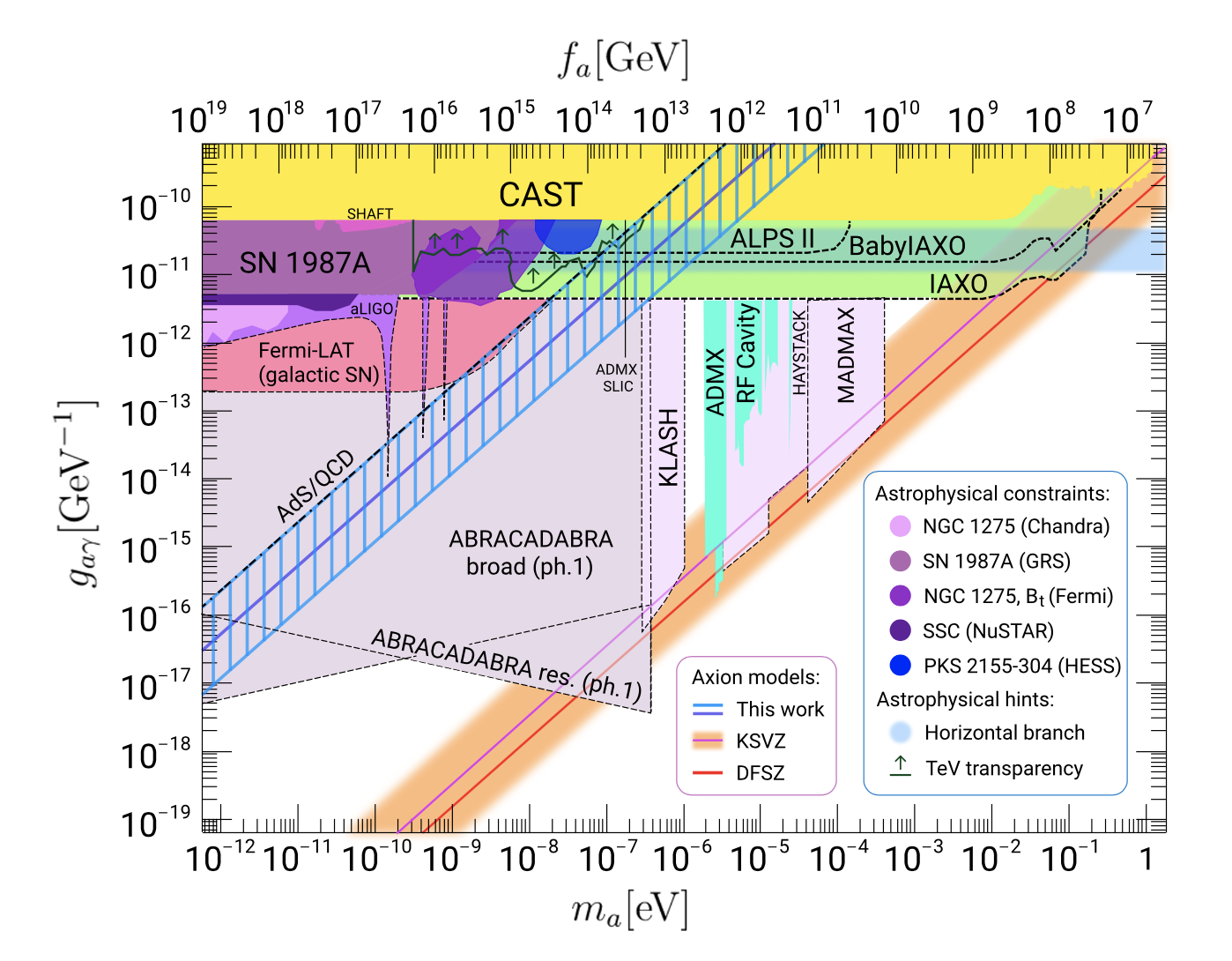}
\caption{Axion-photon coupling as a function of axion mass and decay constant for various axion models. The plot shows the existing and predicted constraints (dotted lines) from experiments together with those derived from astrophysical data and theoretical models. We thank the authors of \cite{Sokolov:2021ydn} for the figure.\\
This article reports on the statuses of the experiments ALPS~II, BabyIAXO and MADMAX.}
\label{Fig:ax-phCoupling}
\end{center}
\end{figure}
For the detection of axions, all three experiments rely on the axion-photon mixing in a background magnetic field (the Sikivie effect \cite{Sikivie:1983ip}) given by the following Lagrangian term:
\begin{equation}
    L_{a\gamma} =  g_{a \gamma } \phi _a \vec{E}\cdot \vec{B}
\end{equation}
where $g_{a \gamma }$ is the axion-photon coupling strength proportional to $1/f_a$ times a model dependent constant, $\phi_a$ is the axion field, $\vec{E}$ is the oscillating electric field of the photon and $\vec{B}$ represents the static magnetic field of the experiments.
Figure \ref{Fig:ax-phCoupling} shows the axion-photon coupling as a function of axion mass, together with existing and predicted constraints from various experiments and astrophysical observations. For reference, the axion-photon couplings in the KSVZ models \cite{Kim:1979if,Shifman:1979if} and in the DFSZ model \cite{Dine:1981rt,Zhitnitsky:1980tq} are compared to a more recent model \cite{Sokolov:2021ydn} indicating the large range of theoretical predictions.

Additional motivation to search for axions and axion-like particles (ALPs, which are not related to the CP conservation in QCD) comes from astrophysical riddles related to the evolution of stars and the propagation of high energy photons in the universe. 
They have been interpreted in various ways, however, a global analysis of all the data indicates a preference for axions and ALPs \cite{Pallathadka:2020vwu,DiLuzio:2021ysg}. 
Interestingly, just one axion with a mass $m_a \approx \SI{e-7}{\electronvolt}$ and $g_{a \gamma } \approx \SI{e-11}{\giga \electronvolt^{-1}}$, as predicted by \cite{Sokolov:2021ydn}, could very well explain the above mentioned riddles, constitute the dark matter and explain the CP conservation in QCD. 
This axion might even be in reach of ALPS~II and BabyIAXO.
In the remainder of this text, we will not strictly differentiate between axions and ALPs anymore.

Basically, axions and other FIPs are searched for by:
\begin{itemize}
    \item Haloscopes looking for axions constituting the cold dark matter halo of our home galaxy;
    \item Helioscopes relying on the Sun as a source of relativistic axions;
    \item Purely laboratory-based experiments not requiring cosmological or astrophysical assumptions.
\end{itemize}

 At DESY, all three techniques are exploited by MADMAX, IAXO and ALPS~II, respectively.
 All are reusing the infrastructure of the former HERA collider\footnote{https://www.desy.de/sites2009/site\_www-desy/content/e409/e69110/e4948/e5101/e5142/e5144/infoboxContent6626/\\
 HERA\_en\_eng.pdf}.
Short summaries of these experiments are given in the following section.
 

\section{Axion searches at DESY}
\subsection{MADMAX}

The MAgnetized Disks and Mirror Axion eXperiment (MADMAX) collaboration \cite{MADMAX:2019pub} lead by the Max-Planck Institute for Physics (Munich, Germany) is developing a new approach to search for axion dark matter in a mass region around $\rm{100\,\mu eV}$ currently not accessible by more traditional approaches based on microwave cavities. 
This mass region is promising as it corresponds to the mass range predicted by post-inflation models \cite{Borsanyi:2016ksw} and the high mass region of pre-inflation models.

MADMAX relies on the conversion of dark matter axions into microwave photons, where the photon energy is given by the axion mass plus an order $\rm{10^{-6}}$ correction due to the dark matter velocity distribution.
The main idea of MADMAX is to exploit the constructive interference of electromagnetic radiation emitted by different surfaces to resonantly enhance the conversion of axions to photons. This is achieved through a series of parallel dielectric disks with a mirror on one side, all within a magnetic field $B$ parallel to the disk surfaces, creating a so-called dielectric haloscope.
\begin{figure}[!ht]
\begin{center}
\includegraphics[scale=0.4]{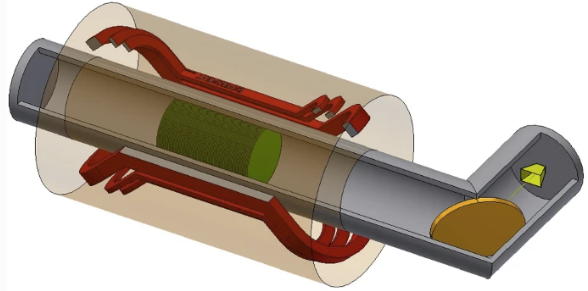}
\caption{Preliminary baseline design of the MADMAX approach (copied from \cite{MADMAX:2019pub}). In the Figure, the three parts of the experiment are shown: (1) magnet (red racetracks); (2) booster, consisting of the mirror (copper disk at the far left) and the dielectric disks (green); (3) the receiver, consisting of the horn antenna (yellow) and the cold preamplifier inside a separated cryostat. The focusing mirror is shown as an orange disk at the right .}
\label{Fig:MADMAX}
\end{center}
\end{figure}
The conceptual sketch of the MADMAX experiment is shown in Figure \ref{Fig:MADMAX}. 
The magnet provides a dipole field of \SI{\sim 9}{\tesla} and an opening of \SI{1.35}{\meter}: dielectric discs with a diameter of more than \SI{1}{\meter} and a mirror reflecting the signal towards the receiver system. 
The distance between the discs can be changed to tune the resonance frequency, which is required to probe different axion masses. 
The booster enhances the axion-induced power by up to four orders of magnitude.
The receiver shall enable the detection of signal photons in the frequency range of \SIrange{10}{100}{\giga \hertz} with a sensitivity of \SI{\sim e-22}{\watt}.

MADMAX is planned to be located in HERA's North Hall \SI{\sim 25}{\meter} below surface. Here, the so-called Cryoplatform will provide the magnet with liquid helium. Furthermore, the iron yoke of the former HERA-experiment H1 will care for the shielding of the magnetic fringe fields and even provide a field strength enhancement at the booster position by about \SI{10}{\percent}.
Initial RF background measurements have demonstrated the suitability of the location.

In the last year, the collaboration has achieved very substantial progress regarding all components of the experiment.
\begin{itemize}
    \item The design of the huge dipole magnet is based on a superconducting cable-in-conduit (CIC). To protect the magnet, a reliable quench-protection system is required in case superconductivity breaks down anywhere in the magnet. In a complex test campaign, CEA/Saclay has clearly demonstrated that the quench propagation velocity within the CIC is fast enough to facilitate the quench detection. Thus, the magnet development has mitigated a crucial potential show-stopper.     
    \item Handling, mounting and positioning of dielectric discs made out of sapphire could be demonstrated. Again, one of the potential show-stoppers could be fully mitigated: a dedicated piezo-motor jointly developed with the company JPE was tested in a strong magnetic field of \SI{5.3}{\tesla} at cryogenic temperatures in vacuum without any issues.
    \item A series of prototype booster tests took place at MPI Munich and CERN and are planned for the future. At CERN the collaboration is using the MORPURGO magnet with a dipole field of up to \SI{1.6}{\tesla} when test beams are shut down. In the year 2022, it was successfully demonstrated that the CERN test beam area allows for physically interesting and competitive axion-like particle dark matter searches. At Munich, a reliable and stably working calibration procedure for a so-called closed-booster-system has been developed. This was extremely important on the path towards a full understanding of the response of the final MADMAX experiment to axion dark matter. 
\end{itemize}

The critical path in the future schedule of MADMAX is mainly given by the availability of funds for the large dipole magnet. Data taking with the final set-up may start in 2030, but already earlier a new prototype magnet could allow for very competitive direct dark matter searches at DESY.

\subsection{BabyIAXO and IAXO}
The International Axion Observatory (IAXO) is a next generation axion helioscope searching for solar axions and ALPs with unprecedented sensitivity. X-ray photons produced via the Sikivie effect (typically in the range of \SIrange{1}{10}{\kilo \electronvolt}) in the magnet bores are focused by high precision X-ray telescopes down to small focal spots at ultra-low background X-ray detectors \cite{IAXO:2019mpb}. The experiment's main goal is to improve the axion-photon coupling sensitivity $g_{a\gamma}$ by more than one order of magnitude with respect to its predecessor experiment CAST \cite{CAST:2007jps}, as illustrated by Figure \ref{Fig:ax-phCoupling}.
IAXO will probe also for the astrophysical hints and search for axions in the eV mass range, which is not accessible in any other axion experiment. 
Furthermore, the physics program is complemented with the possibility to probe different axion production models in the Sun by investigating the axion-electron $g_{ae}$ and the axion-nucleon $g_{an}$ couplings  \cite{Jaeckel:2018mbn,DiLuzio:2021qct}.
At a later point in time, the IAXO magnet could be used to search for dark matter halo axions by the accommodation of additional equipment, like microwave antennas and cavities similar to the RADES project at CERN \cite{IAXO:2019mpb}. 

\begin{figure}[!ht]
\begin{center}
\includegraphics[scale=0.4]{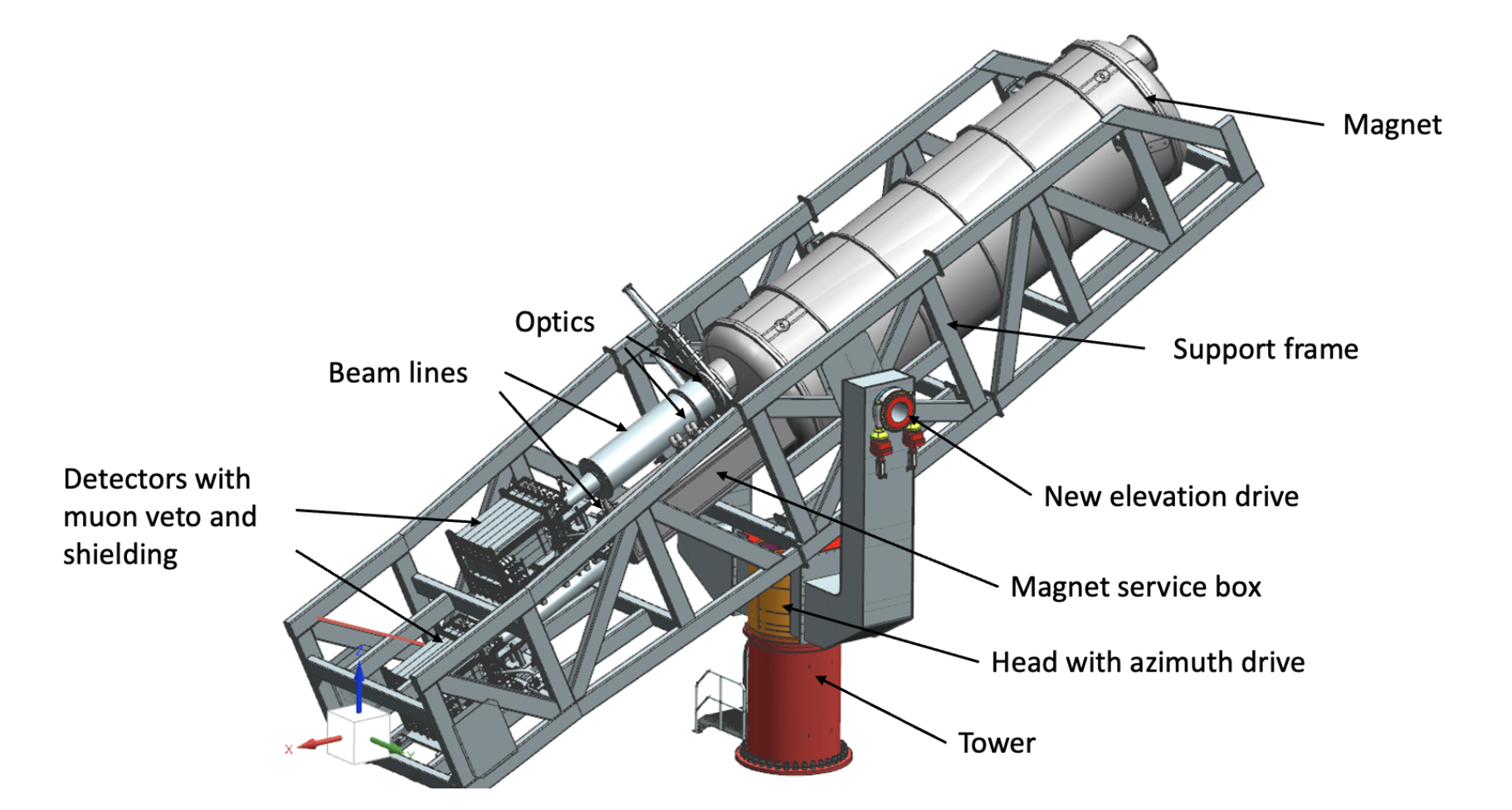}
\caption{Overview of the BabyIAXO experiment. The system features a total length of \SI{\sim 21}{\meter} \cite{IAXO:2020wwp}.}
\label{Fig:BabyIAXO_CDR}
\end{center}
\end{figure}

The first step towards the full IAXO will be the BabyIAXO experiment.
It is conceived with two main objectives: First, to serve as a full technological prototype for all subsystems of IAXO to prove full system integration and hence mitigate risks. Second, to operate as a fully fledged helioscope with own potential for discovery, e.g. by exceeding CAST sensitivity on $g_{a\gamma}$ by a factor $\sim 4$ in the same axion mass range. BabyIAXO will be based on a \SI{10}{\meter} long superconducting magnet (\SI{\sim 2}{\tesla}) with two bores, each with a diameter of \SI{70}{\centi \meter}. As depicted by Figure \ref{Fig:BabyIAXO_CDR}, the two equipped individual detection lines will feature an X-ray optics and an ultra-low background X-ray detector each, with comparable parameters and dimensions as foreseen for IAXO \cite{IAXO:2020wwp}. 

The BabyIAXO magnet relies on a common coil design based on two flat racetrack-coils with counter-flowing currents providing a dipole field in both bores. Superconducting Al-stabilised Rutherford cables will allow for a challenging dry detector magnet concept achieved by cryocoolers and He-gas circulators. However, since the Russian invasion of Ukraine in February 2022, collaboration with Russian institutes is frozen and the collaboration cannot get the magnet cable from Russia as planned before.  
No vendor for such cables, produced with co-extrusion technology, is available in non-Russian industry anymore. Due to the fact that this problem concerns many other particle physics experiments as well, a global effort to find sophisticated alternatives has been initiated and coordinated by CERN, resulting in a dedicated workshop in September 2022.
Different possible solutions have been identified and are being followed up. 
It is hoped that by summer 2023,   updated time schedules and cost estimates for the BabyIAXO magnet can be provided.

With respect to the X-ray optics, BabyIAXO will use one XMM-Newton flight spare module on loan by ESA in one magnetic bore, while the other bore will be equipped with a custom-made and newly built X-ray optics module. To detect the X-rays with high sensitivity, a variety of detector technologies are considered, mainly divided into \textit{discovery} and \textit{energy resolving} detectors \cite{IAXO:2020wwp}. The baseline option relies on small gas chambers (typically \SI{3}{\centi \meter} thick and \SI{6}{\centi \meter} wide) read out by a finely segmented micro-mesh gas structures (Micromegas).
Similar detectors were already successfully operated in CAST. While the other detector technologies are in the R\&D phase, the BabyIAXO Micromegas detector prototypes are currently undergoing final tests in the low background environments of the laboratories at Canfranc. In addition, special efforts and studies are conducted with respect to radio-pure materials, high efficiency veto systems and dedicated shielding with a special focus on neutron backgrounds. Lastly, the main design of the structure and drive system, which re-uses a modified tower and positioner of a Cherenkov Telescope Array (CTA) prototype (from DESY in Zeuthen), is close to being finished and ready for production.

After the approval of the experiment to be hosted at DESY, the collaboration has already taken first steps towards its construction. Currently, the first data taking with the full BabyIAXO experiment is foreseen for 2028. However, the future schedule is mainly driven by the status and availability of the superconducting cable for the magnet.

\subsection{ALPS~II}
The Any Light Particle Search (ALPS) number 2 is a light-shining-through-a-wall (LSW) experiment \cite{Bahre:2013ywa}. 
It will improve sensitivity on the axion-photon coupling by a factor of $10^3$ compared to its predecessors. 
This jump in sensitivity will be achieved by a long string of superconducting dipole magnets and two mode-matched 
optical cavities before and after the light-tight wall as proposed for the first time more than 30 years ago \cite{Hoogeveen:1990vq}. The presence of these cavities will resonantly increase the probability of the production of axions and the probability of their re-conversion into photons.
For the first time, ALPS~II will allow probing for axions in a model-independent fashion beyond present-day limits from astrophysics. 
\begin{figure}[!ht]
\begin{center}
\includegraphics[scale=0.35]{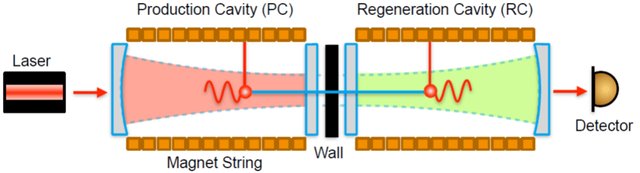}
\includegraphics[scale=0.3]{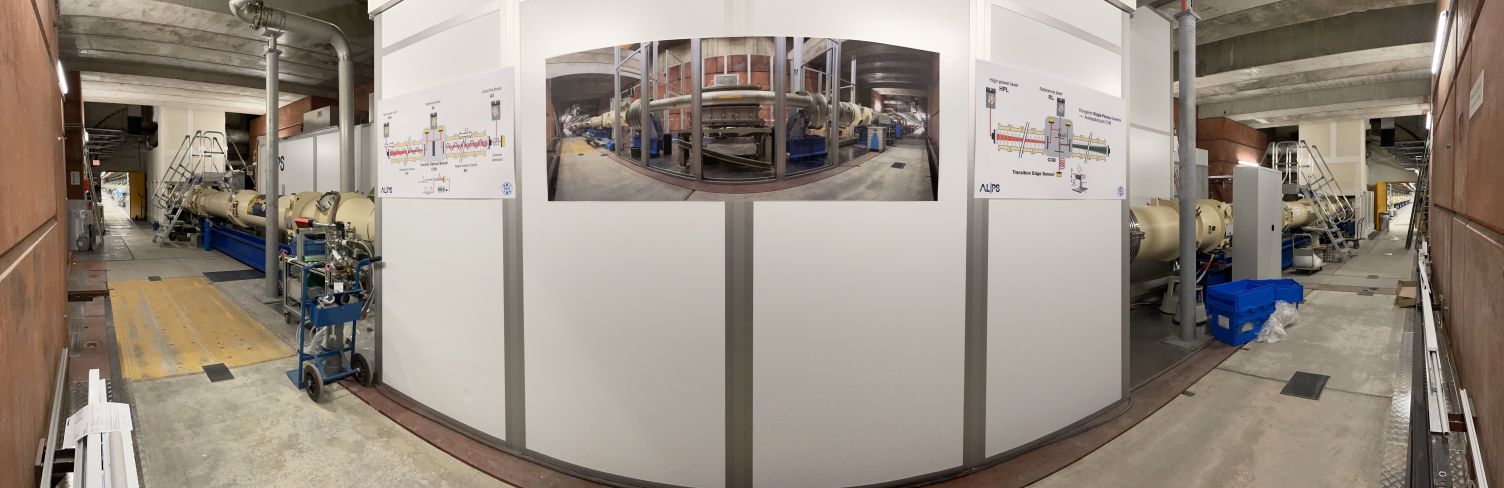}
\caption{Schematic layout of the ALPS~II experiment \cite{Albrecht:2020ntd}, left, and a panoramic picture of the installation in the straight HERA tunnel section around the North hall (right).}
\label{Fig:ALPSII}
\end{center}
\end{figure}

ALPS~II, see Figure \ref{Fig:ALPSII}, consists of two \SI{122}{\meter} long high-finesse optical cavities \cite{Ortiz:2020tgs} inside two strings of 12 superconducting HERA dipole magnets each \cite{Albrecht:2020ntd}. 
The probability for light converting to axions (which easily pass any barrier) and axions converting back to light is given by (for axion masses below \SI{0.1}{\milli \electronvolt})
\begin{equation}
    P_{\gamma \to \, a \, \to \gamma} = 
    \frac{1}{16} \beta_{PC} \beta_{RC} (g_{a\gamma \gamma} B L)^4 = 
    6 \times 10^{-38 } \beta{F}_{PC} \beta{F}_{RC} \times 
    \left(\frac{g_{a\gamma \gamma}}{10^{-10} \rm{~GeV^{-1}}} \frac{B}{1 \rm{~T}} \frac{L}{10 \rm{~m}}\right)^4
\end{equation}
resulting in $10^{-25}$ for the ALPS~II parameters $\beta_{PC}=5000$, $\beta_{RC}= 40000$, $B = \SI{5.3}{\tesla}$, $L = \SI{105.6}{\meter}$ and $g_{a\gamma \gamma} = \SI{2e-11}{\giga \electronvolt^{-1}}$ (motivated by astrophysics).
Thus, with \SI{30}{\watt} of \SI{1064}{\nano \meter} photons injected in the PC, about $2$ photons/day behind the wall are expected.

ALPS~II will exploit two independent signal detection systems, each with very different systematic uncertainties. 
This approach will help to increase confidence that any signals observed with the same intensity in both detectors are indeed the result of a photon-axion conversion-re-conversion process. Since the two detectors require different optical systems to operate, they cannot be used in parallel. 
The first scheme to be implemented will be a heterodyne detection (HET) scheme \cite{Hallal:2020ibe,Spector:2023nap}, which measures the interference beat note between a laser, called a local oscillator (LO), and the regenerated photon field. 
The second detection scheme will use a Transition Edge Sensor (TES) \cite{Shah:2021wsp,Dreyling-Eschweiler:2015pja} operated at about \SI{100}{\milli \kelvin}. It allows for counting individual \SI{1064}{\nano \meter} photons with an energy resolution of \SI{\sim 7}{\percent}.

The installation of ALPS~II began in 2019.
In March 2022 the magnet string was successfully tested and in September 2022 the optics installation was completed for the initial science run.
The experiment is now close to start operation. 

This run based on the HET detection scheme will happen in the first months of 2023. 
It will not include the production cavity before the wall to optimise for the study of stray light, but already go  by a factor of 100 in the axion-photon coupling beyond earlier LSW experiments.
The full optical system will be used in the second half of 2023 and a HET science run with upgraded optics is planned for 2024. The further scheduling depends on the outcome of the HET science run, results of ongoing R\&D, resources and worldwide science advancements. The future program might include a TES-based science run, vacuum magnetic birefringence measurements, FIP searches with optimized optics and/or extension of the ALP mass reach and a dedicated search for high-frequency gravitational waves.

\section{Conclusion}
DESY in Hamburg is planning for three larger scale axion  experiments exploiting the LSW technique, solar axion and axion dark matter searches, all strongly pushed for by international collaborations.
The first one, ALPS~II, will start its science program soon. BabyIAXO is ready to launch construction when a new road-map for realizing the magnet exists and the funding is clarified; MADMAX is in the prototyping phase. 
\begin{figure}[!bp]
  \centering
  \subfloat[]{\includegraphics[width=0.52\textwidth]{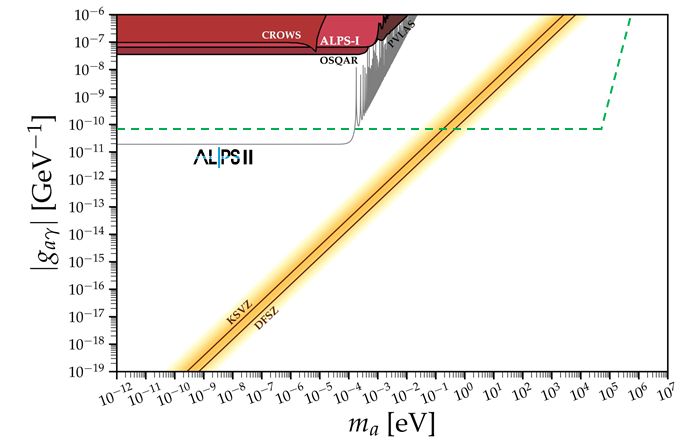}\label{Fig:ALPSIIlim}}
  \subfloat[]{\includegraphics[width=0.48\textwidth]{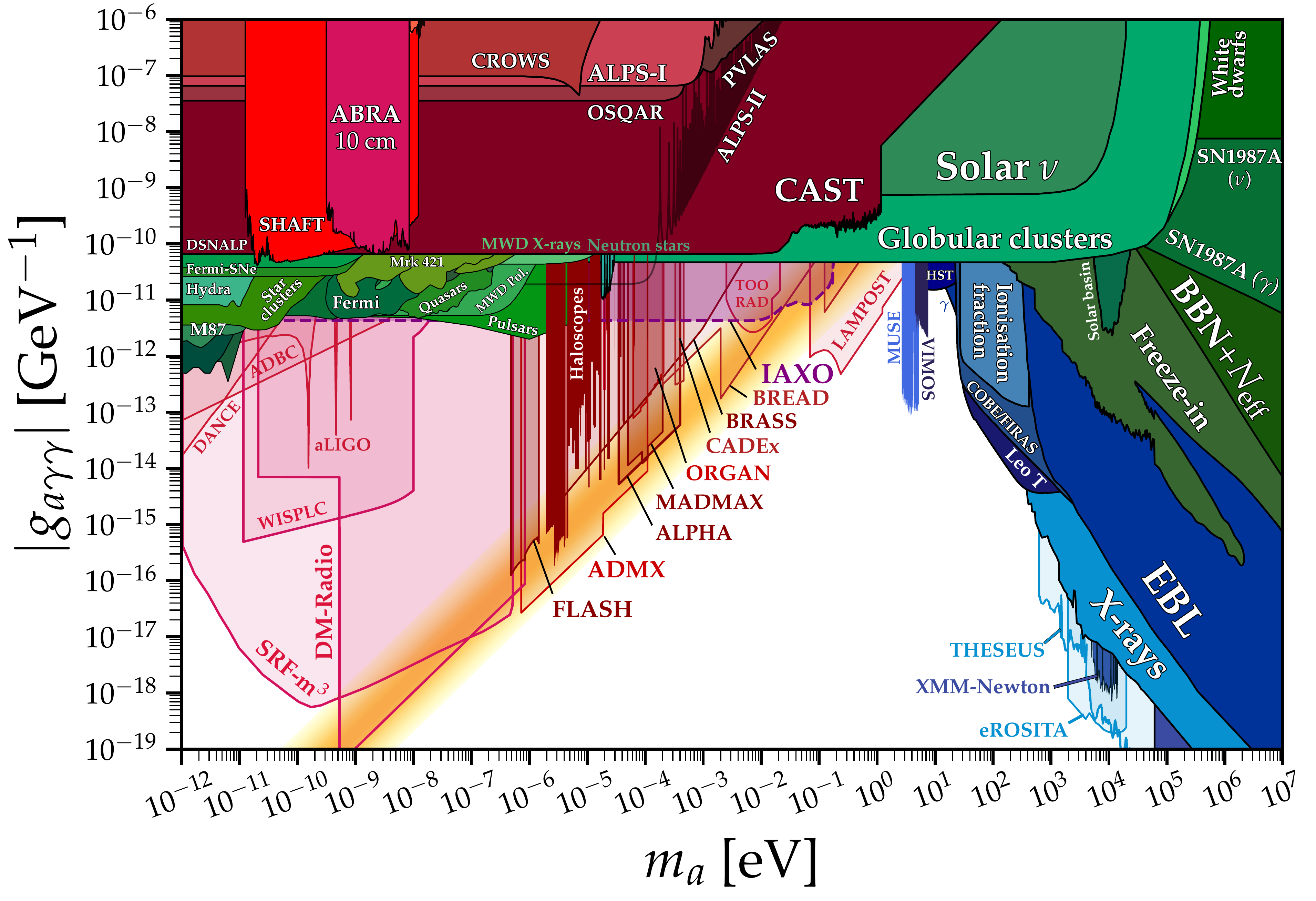}\label{Fig:ALPSIIlim2}}
  \caption{(a) ALP parameter space for model independent searches experiments \cite{AxionLimits}, the dashed line indicates limits from astrophysics. (b) Overview of results, prospects, and hints in the axion/ALP parameter space \cite{AxionLimits}.}
  \label{Fig:Lim_}
\end{figure}
Figure \ref{Fig:Lim_} shows the projection of the sensitivities of all three experiments together with existing limits.
It is very obvious that this unique set of axion experiments in the old HERA premises at DESY offers a likewise unique discovery potential to solve major particle physics, astrophysics and cosmological questions.

\bibliographystyle{unsrt}
\bibliography{bib_ALPS.bib}
\end{document}